\documentclass{iau}

\usepackage{amsmath}
\usepackage{graphicx}
\usepackage{multirow}
\usepackage{chemformula}
\usepackage{xspace}
\usepackage{upgreek}
\let\ce\ch
\newcommand{\kms}{\,km\,s$^{-1}$\xspace} 
\newcommand{\pc}{\,pc\xspace} 
\newcommand{\K}{\,K\xspace} 
\newcommand{\GHz}{\,GHz\xspace} 
\newcommand{\mm}{\,mm\xspace} 

\begin{document}

\lefttitle{\textit{San Andr{\'e}s et al.}}
\righttitle{The Galactic Centre G+0.633-0.0604 Molecular Cloud: A New Gold mine for Astrochemistry}

\jnlPage{1}{7}
\jnlDoiYr{2021}
\doival{10.1017/xxxxx}

\aopheadtitle{Proceedings IAU Symposium}
\editors{M. Zaja\v{c}ek,  T. Je\v{r}\'{a}bkov\'{a}, V. Karas, R. Schödel \&  P. Sukov\'{a}, eds.}

\title{The Galactic Centre G+0.633-0.0604 Molecular Cloud: A New Gold Mine for Astrochemistry}

\author{David San Andr{\'e}s$^1$, V{\'i}ctor M. Rivilla$^1$, Laura Colzi$^1$, Miguel Sanz-Novo$^2$, Sergio Mart{\'i}n$^{3,4}$, Izaskun Jim{\'e}nez-Serra$^1$, and Shaoshan Zeng$^5$}

\affiliation{$^1$ Centro de Astrobiolog{\'i}a (CAB), CSIC-INTA, Carretera de Ajalvir km 4, 28850 Torrej{\'o}n de Ardoz, Madrid, Spain ; email: 
\email{david.sanandres@cab.inta-csic.es}\\[3pt]
$^2$ Center for Astrochemical Studies, Max-Planck-Institut f{\"{u}}r extraterrestrische Physik, Giessenbachstrasse 1, Garching bei Munchen, 85748, Germany\\[3pt]
$^3$ European Southern Observatory, Alonso de C{\'o}rdova, 3107, Vitacura, Santiago 763-0355, Chile\\[3pt]
$^4$ Joint ALMA Observatory, Alonso de C{\'o}rdova, 3107, Vitacura, Santiago 763-0355, Chile\\[3pt]
$^5$ Star and Planet Formation Laboratory, Pioneering Research Institute (PRI), RIKEN, 2-1 Hirosawa, Wako, Saitama, 351-0198, Japan}

\begin{abstract}
Astrochemistry is living a golden age, with more than a quarter of the $\sim$350 molecules in the current interstellar census having been detected over the last three years. One of the sources driving this progress is the G+0.693-0.027 cloud, located in the northern part of the Galactic Centre Sgr B2 complex. In this contribution, we present the astrochemical characterisation of G+0.633-0.0604, a newly discovered chemically rich molecular cloud at the southern edge of Sgr B2. With an inventory of $>$120 species, G+0.633 provides robust second detections of several prebiotic molecules only reported towards G+0.693, establishing it as the first confirmed astrochemical twin of G+0.693 while demonstrating that the extraordinary chemistry of this cloud is not unique. Furthermore, G+0.633 offers an observational advantage over G+0.693 since it displays half narrower linewidths. Together, G+0.633 and G+0.693 form a unique benchmark pair for unveiling molecular complexity and prebiotic chemistry in the interstellar medium.
\end{abstract}

\begin{keywords}
Astrochemistry, Galaxy:center, ISM:clouds, ISM:molecules
\end{keywords}

\maketitle

\section{Introduction}

The detection of molecules beyond the Earth marked a major milestone in the history of astronomy. The very first evidence for interstellar molecules came nearly a century ago with the identification of the simple diatomic species \ce{CN}, \ce{CH}, and \ce{CH+} through optical absorption lines (\citealt{Dunham1937}; \citealt{Swings&Rosenfeld1937}; \citealt{McKellar1940}; \citealt{Douglas&Herzberg1941}). Although these pioneering discoveries demonstrated that molecules are also present in the interstellar medium (ISM), they remained isolated detections for nearly two decades. The advent of radio astronomy completely transformed this picture. Following the first radio detection of the \ce{OH} radical by \citet{Weinreb1963} and, shortly afterwards, ammonia (\ce{NH3}) towards the centre of our Galaxy \citep{Cheung1968}, radio observations became the primary avenue for unveiling the molecular Universe. Continuous advances in radio astronomy, driven by the successive generations of radio telescopes, increasingly sophisticated receivers, and the improvement in laboratory spectroscopy and spectral analysis techniques, have enabled an extraordinary expansion of the known interstellar molecular inventory. Today, nearly 350 molecular species have been identified in the ISM\footnote{\url{https://cdms.astro.uni-koeln.de/classic/molecules}}, spanning simple diatomic radicals to complex organic molecules (COMs; denoting carbon-based molecules with at least six atoms; \citealt{Herbst2020}) containing more than a dozen atoms. Remarkably, nearly one third of these species have been discovered just in the last three years, highlighting the rapid pace at which astrochemistry is currently evolving. 

Today, interstellar molecules are known to populate virtually every astrophysical environment, encompassing sources with widely different physical conditions and evolutionary stages, from the coldest quiescent dark clouds to the hottest and most extreme environments in the Galaxy, and extending well beyond the Milky Way into other galaxies (see, e.g., \citealt{Jimenez-Serra2025, Martin2021}). As such, molecules have become powerful tracers of the physical and chemical processes governing the interstellar medium. Nevertheless, the discovery of increasingly complex molecules remains concentrated towards only a handful of exceptionally chemically rich sources, making the search for new molecular reservoirs a key step towards understanding the chemistry that drives molecular complexity in the ISM. Among these privileged laboratories, the Sgr B2 giant molecular cloud complex, located in the Central Molecular Zone (CMZ) of our Galaxy at $\sim$100\pc east of the Sgr A$^*$ supermassive central black hole, particularly stands out. It emerges as one of the most massive and active star-forming sites in the Galaxy, delineated from north to south by three prominent hot cores (N, M, and S; see Fig.~\ref{fig:summary_figure}), thus offering an excellent window to explore the astrochemical input for star formation. Among them, Sgr B2(N) is particularly chemically rich, and pioneer for many years in advancing the search for new molecular species (e.g., \citealt{Belloche2013, Belloche2020}). Nevertheless, when it comes to molecular richness, a nearby but physically distinct source has attracted special attention over the last years: the G+0.693-0.027 molecular cloud.

\subsection{The G+0.693-0.027 molecular cloud: a benchmark astrochemical laboratory}

Located in the northern part of the Sgr B2 complex (see Fig.~\ref{fig:summary_figure}), the G+0.693-0.027 molecular cloud (hereafter G+0.693) has become one of the richest molecular reservoirs known in the ISM. Despite its close proximity (just $\sim$55$^{\prime\prime}$) to the active star-forming core Sgr B2(N), G+0.693 shows no evidence of ongoing star formation, but instead exhibits a unique combination of physical and chemical properties that make it 
an exceptional astrochemical laboratory. First, G+0.693 is characterised by moderate gas kinetic temperatures, $T_\text{kin}$, of $\sim$70$-$140\K \citep{Zeng2018, Colzi2024}, which sharply contrast with the much lower excitation temperatures, $T_\text{ex}$, measured for most molecules of $\sim$3$-$20\K (e.g., \citealt{Colzi2022, SanAndres2023}), leading to widespread subthermal excitation. As a result, the molecular spectra are significantly cleaner than those of the Sgr B2 hot cores, substantially reducing line blending and lowering the line confusion limit, thereby greatly facilitating the identification of weak molecular transitions. Second, G+0.693 is believed to be permeated by low-velocity shocks generated by the large-scale cloud-cloud collision thought to be shaping the entire Sgr B2 complex, as evidenced by the enhanced \ce{HNCO} emission, a well-established shock tracer, coincident with the position and systemic velocity of the cloud (see Fig.~\ref{fig:summary_figure}). These shocks efficiently sputter the icy mantles of dust grains releasing into the gas phase a large variety of molecules formed there \citep{Martin2008, Zeng2020}, thereby significantly increasing their gas-phase abundances and, consequently, enhancing their detectability.

The combination of widespread subthermal excitation and large-scale shock processing has made G+0.693 one of the most productive astrochemical laboratories in the Galaxy. Around 40\% of all molecular species currently reported in the ISM have been already identified towards this cloud, including numerous COMs and many species of prebiotic relevance, several of which have so far only been detected towards it (e.g., \citealt{Rivilla2020, Rivilla2021, Rivilla2023, Sanz-Novo2023, Sanz-Novo2026, SanAndres2024}). 
These discoveries have firmly established G+0.693 as one of the benchmark astrochemical laboratories in the Galaxy. Nevertheless, whether such extraordinary chemical complexity is unique or instead a common outcome of shock-driven chemistry has remained an open question.

\section{The G+0.633-0.0604 molecular cloud: a new gold mine for astrochemistry}

\begin{figure}
    \centering
    \includegraphics[width=\linewidth]{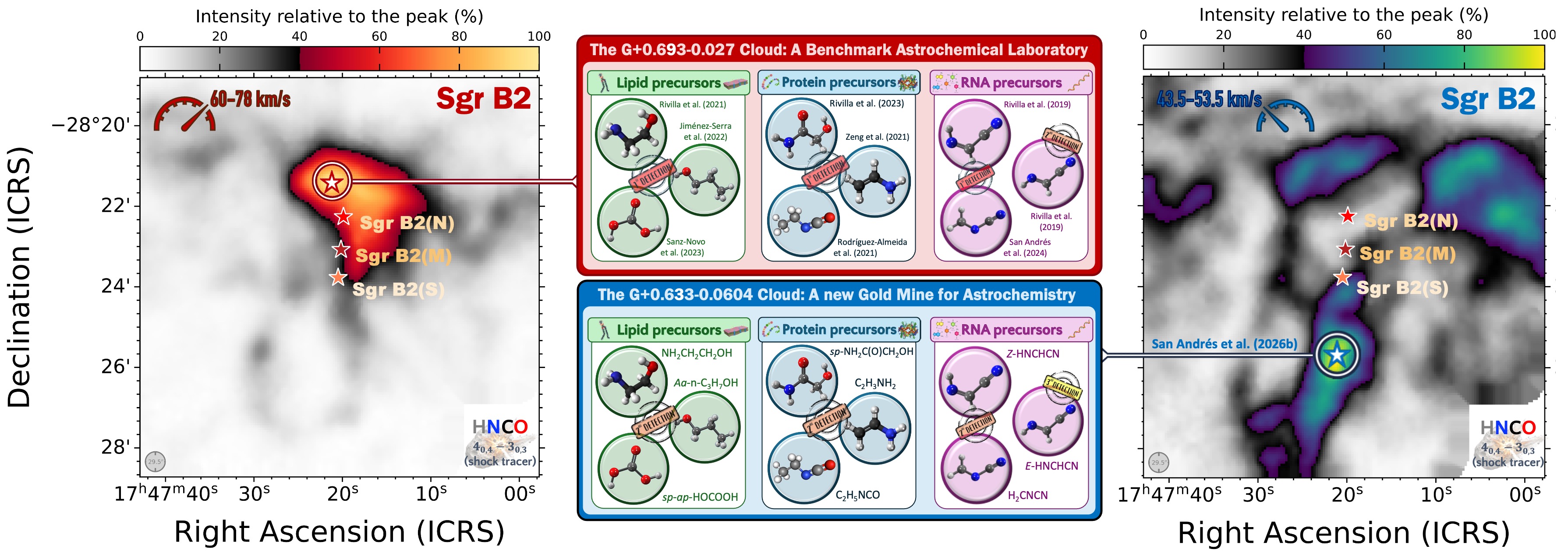}
    \caption{Overview of the two astrochemical laboratories presented in this contribution. The left and right panels show the integrated intensity (moment 0) maps of the \ce{HNCO} $4_{0{,}4}-3_{0{,}3}$ emission within the Sgr B2 complex computed at the velocities of G+0.693 (left panel in red) and G+0.633 (right panel in blue), respectively, illustrating that both clouds emerge right at the most prominent \ce{HNCO} emission peaks. The central panels summarise representative examples of molecules of prebiotic relevance first detected towards G+0.693 (see references) and now identified in G+0.633 (see \citealt{SanAndres2026_G0633-II}), grouped according to their potential connection with major biochemical building blocks (lipid, protein, and RNA precursors). The striking correspondence between the molecular inventories of these two clouds highlights the G+0.633 as the first confirmed astrochemical twin of G+0.693.}
    \label{fig:summary_figure}
\end{figure}

The recent identification of the G+0.633-0.0604 molecular cloud (hereafter G+0.633), physically characterised in \citet{SanAndres2026_G0633-I}, has provided the first compelling evidence that the extraordinary chemistry of G+0.693 is not unique. In fact, this new cloud shares many of the characteristics that made G+0.693 unique. Located at the southern edge of the Sgr B2 complex (Fig.~\ref{fig:summary_figure}), approximately 10\pc in projected distance south of G+0.693, this new cloud also emerges at the peak of \ce{HNCO} emission at lower velocities ($\sim$48.5\kms), indicating a shock-dominated environment analogous to that of its northern counterpart ($\sim$69\kms). Moreover, G+0.633 exhibits similarly low $T_\text{ex}$ of $\sim$3$-$20\K compared to the much higher $T_\text{kin}$ found of $\sim$55$-$90\K, revealing that the molecular emission in G+0.633 is likewise subthermal \citep{SanAndres2026_G0633-I, SanAndres2026_G0633-II}. This combination immediately identified G+0.633 as a particularly promising candidate to harbour a molecular richness comparable to that of G+0.693.

Its astrochemical characterisation, presented in \citet{SanAndres2026_G0633-II}, builds upon the ongoing broadband molecular line survey introduced in \citet{SanAndres2026_G0633-I}, covering nearly 100\GHz of aggregated bandwidth across the 7\mm, 3\mm and 1.3\mm windows through observations with the Yebes 40m (Guadalajara, Spain), IRAM 30m (Granada, Spain) and APEX (Chajnantor, Chile) radio telescopes, and achieving sub-mK sensitivity. To date, this survey has revealed over 120 molecular species containing all six essential biogenic elements (\ce{C}, \ce{H}, \ce{O}, \ce{N}, \ce{P}, and \ce{S}), including the aromatic cycle benzonitrile (c-\ce{C6H5CN}; \citealt{Rivilla2026}), immediately placing G+0.633 among the richest molecular reservoirs in the Galaxy.

Within its molecular inventory, species of prebiotic relevance deserve special attention, many of which had previously been identified only towards G+0.693. Here, we highlight some of the most representative examples (see Fig.~\ref{fig:summary_figure}). Within the family of lipid precursors, G+0.633 hosts the second interstellar detection of ethanolamine (\ce{NH2CH2CH2OH}, first detected in G+0.693 by \citet{Rivilla2021}), one of the most common constituents of the hydrophilic head groups of phospholipids, the key units that form modern biological membranes. Moving to protein precursors, G+0.633 provides the second unambiguous interstellar detection of $sp$-glycolamide ($sp$-\ce{NH2C(O)CH2OH}, first detected in G+0.693 by \citealt{Rivilla2023}), a structural isomer of glycine, one of the simplest amino acids. Finally, within the framework of the RNA world hypothesis, G+0.633 hosts all three cyanomethanimine isomers: $Z$- and $E$-$C$-cyanomethanine (\ce{HNCHCN}, reported towards G+0.693 by \citealt{Rivilla2019}) together with $N$-cyanomethanimine (\ce{H2CNCN}, previously known only towards G+0.693; \citealt{SanAndres2024}), regarded as key intermediates in proposed synthetic pathways leading to adenine, one of the purine nucleobases present in both RNA and DNA chains. These detections, together with the remarkable similarity of the overall molecular inventory \citep{SanAndres2026_G0633-II}, establish G+0.633 as the first confirmed astrochemical twin of G+0.693.

\begin{figure}
    \centering
    \includegraphics[width=\linewidth]{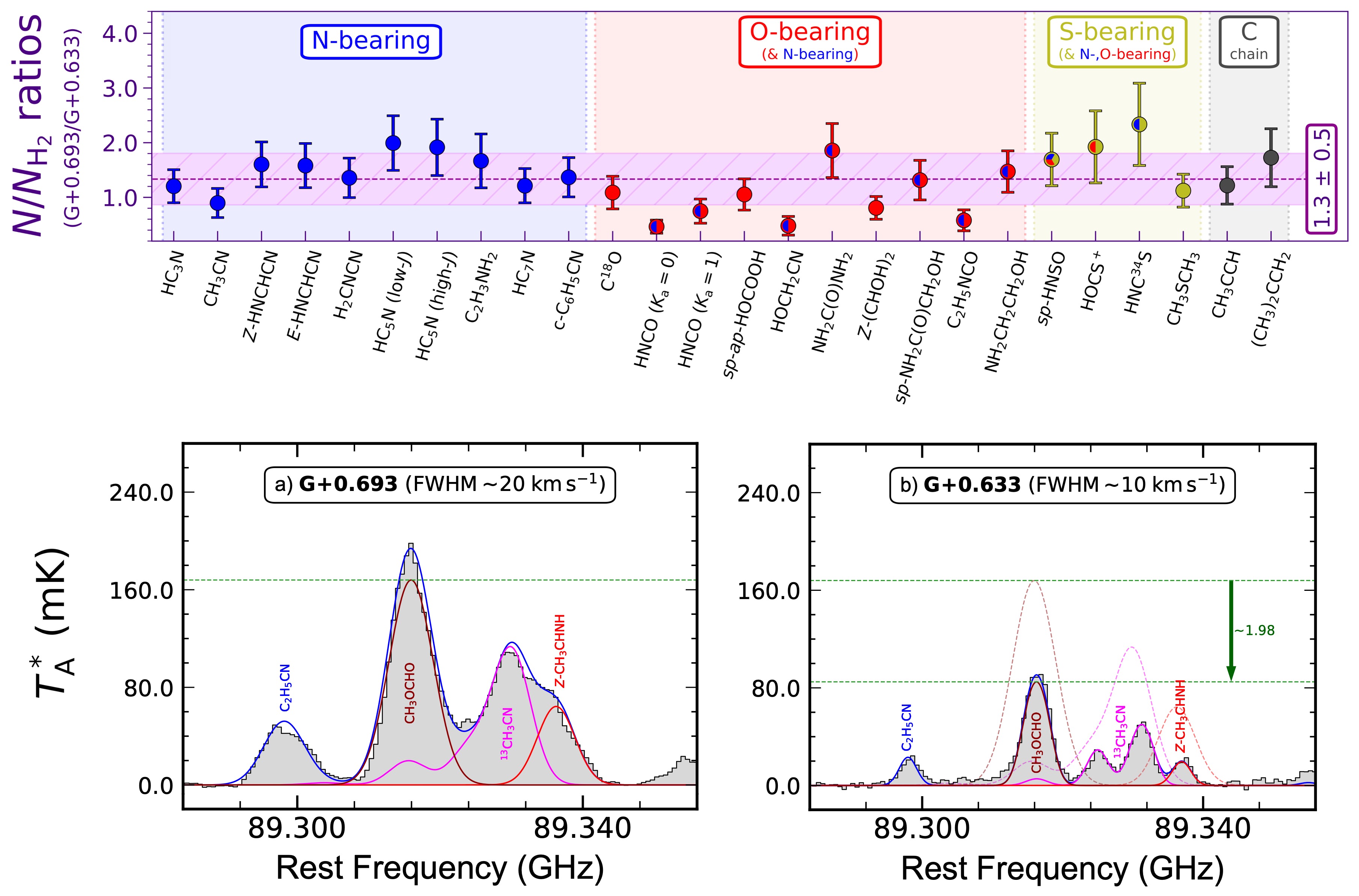}
    \caption{Comparison of the molecular properties between G+0.693 and G+0.633. \textit{Top}: relative abundance ratios ($N/N_{\ce{H2}}$) of G+0.693 over G+0.633 for all \ce{N}-, \ce{O}-, \ce{S}-bearing and \ce{C} chain species reported in G+0.633 so far \citep{Rivilla2026, SanAndres2026_G0633-I, SanAndres2026_G0633-II}. The dashed violet line and grated region represent the average value and the 1$\upsigma$ uncertainty interval (see right box), respectively. \textit{Bottom}: comparison of the widths and relative intensities (dashed green lines, arrow and label) for a set of lines in G+0.693 (left) versus G+0.633 (right). The black histograms indicate the observed spectrum, the solid red, brown, and magenta lines outline the particular species contributing to the set (see labels), while the solid blue line delineates their combined profiles and the emission from the rest of molecules characterised in these two clouds.}
    \label{fig:G0693-G0633_comparison}
\end{figure}

In addition to the nearly identical molecular inventories shared by both clouds, the strongest evidence that G+0.633 is the first confirmed astrochemical twin of G+0.693 lies in the tight correlations in molecular abundances across a wide range of chemical families (see Fig.~\ref{fig:G0693-G0633_comparison}). This remarkable agreement indicates that both sources have experienced similar chemical histories and suggests that their extraordinary molecular richness is not the result of exceptional local conditions, but rather the natural outcome of a common shock-driven chemistry operating under similar physical conditions \citep{SanAndres2026_G0633-I}. Consequently, G+0.633 and G+0.693 emerge as a powerful benchmark pair for astrochemical studies, suitable not only for confirming molecular detections, but also for validating and refining astrochemical models.

Besides its scientific implications, G+0.633 also offers a clear observational advantage over its northern counterpart. The linewidths of molecular emission lines are approximately a factor of two narrower than those measured towards G+0.693 ($\sim$10\kms versus $\sim$20\kms; see Fig.~\ref{fig:G0693-G0633_comparison}), substantially reducing spectral blending and enabling weaker molecular transitions to be identified with greater confidence. This makes G+0.633 an exceptionally favourable laboratory for extending the census of increasingly complex interstellar molecules and probing the astrochemical pathways leading to molecular complexity in the ISM.

\section{Conclusions}

The astrochemical characterisation of the newly identified G+0.633-0.0604 molecular cloud demonstrates that the extraordinary molecular richness previously associated with G+0.693 is not unique within the Galactic Centre. With a molecular inventory already exceeding 120 identified species, including robust second interstellar detections of several prebiotically relevant molecules, G+0.633 emerges as the first confirmed astrochemical twin of G+0.693 and joins it among the richest molecular reservoirs currently known in the ISM.

The remarkable correspondence between the molecular inventories and abundances of both clouds indicates that the extraordinary chemical richness of these sources is a natural outcome of shock-driven chemistry rather than an exceptional case. Together, G+0.633 and G+0.693 establish a unique benchmark pair for astrochemical studies, while the significantly narrower molecular linewidths of G+0.633 make it an exceptionally favourable laboratory for extending the census of increasingly complex interstellar molecules and further advancing our understanding of the chemical pathways leading to molecular complexity in the ISM.

\bibliography{biblio}

@ARTICLE{Belloche2013,
       author = {{Belloche}, A. and {M{\"u}ller}, H.~S.~P. and {Menten}, K.~M. and {Schilke}, P. and {Comito}, C.},
        title = "{Complex organic molecules in the interstellar medium: IRAM 30 m line survey of Sagittarius B2(N) and (M)}",
      journal = {\aap},
         year = 2013,
        month = nov,
       volume = {559},
          eid = {A47},
        pages = {A47},
          doi = {10.1051/0004-6361/201321096},
archivePrefix = {arXiv},
       eprint = {1308.5062},
 primaryClass = {astro-ph.GA},
       adsurl = {https://ui.adsabs.harvard.edu/abs/2013A&A...559A..47B}
}

@ARTICLE{Belloche2020,
       author = {{Belloche}, A. and {Garrod}, R.~T. and {M{\"u}ller}, H.~S.~P. and {Menten}, K.~M. and {Medvedev}, I. and {Thomas}, J. and {Kisiel}, Z.},
        title = "{Re-exploring Molecular Complexity with ALMA (ReMoCA): interstellar detection of urea}",
      journal = {\aap},
         year = 2019,
        month = aug,
       volume = {628},
          eid = {A10},
        pages = {A10},
          doi = {10.1051/0004-6361/201935428},
archivePrefix = {arXiv},
       eprint = {1906.04614},
 primaryClass = {astro-ph.GA},
       adsurl = {https://ui.adsabs.harvard.edu/abs/2019A&A...628A..10B}
}

@ARTICLE{Cheung1968,
       author = {{Cheung}, A.~C. and {Rank}, D.~M. and {Townes}, C.~H. and {Thornton}, D.~D. and {Welch}, W.~J.},
        title = "{Detection of NH$_{3}$ Molecules in the Interstellar Medium by Their Microwave Emission}",
      journal = {Phys.~Rev.~Lett.},
         year = 1968,
        month = dec,
       volume = {21},
       number = {25},
        pages = {1701-1705},
          doi = {10.1103/PhysRevLett.21.1701},
       adsurl = {https://ui.adsabs.harvard.edu/abs/1968PhRvL..21.1701C}
}

@ARTICLE{Colzi2022,
       author = {{Colzi}, Laura and {Mart{\'\i}n-Pintado}, Jes{\'u}s and {Rivilla}, V{\'\i}ctor M. and {Jim{\'e}nez-Serra}, Izaskun and {Zeng}, Shaoshan and {Rodr{\'\i}guez-Almeida}, Lucas F. and {Rico-Villas}, Fernando and {Mart{\'\i}n}, Sergio and {Requena-Torres}, Miguel A.},
        title = "{Deuterium Fractionation as a Multiphase Component Tracer in the Galactic Center}",
      journal = {\apjl},
         year = 2022,
        month = feb,
       volume = {926},
       number = {2},
          eid = {L22},
        pages = {L22},
          doi = {10.3847/2041-8213/ac52ac},
archivePrefix = {arXiv},
       eprint = {2202.04111},
 primaryClass = {astro-ph.GA},
       adsurl = {https://ui.adsabs.harvard.edu/abs/2022ApJ...926L..22C}
}

@ARTICLE{Colzi2024,
       author = {{Colzi}, L. and {Mart{\'\i}n-Pintado}, J. and {Zeng}, S. and {Jim{\'e}nez-Serra}, I. and {Rivilla}, V.~M. and {Sanz-Novo}, M. and {Mart{\'\i}n}, S. and {Zhang}, Q. and {Lu}, X.},
        title = "{Excitation and spatial study of a prestellar cluster towards G+0.693-0.027 in the Galactic centre}",
      journal = {\aap},
         year = 2024,
        month = oct,
       volume = {690},
          eid = {A121},
        pages = {A121},
          doi = {10.1051/0004-6361/202451382},
archivePrefix = {arXiv},
       eprint = {2408.17141},
 primaryClass = {astro-ph.GA},
       adsurl = {https://ui.adsabs.harvard.edu/abs/2024A&A...690A.121C}
}

@ARTICLE{Douglas&Herzberg1941,
       author = {{Douglas}, A.~E. and {Herzberg}, G.},
        title = "{Note on CH\^\{+\} in Interstellar Space and in the Laboratory.}",
      journal = {\apj},
         year = 1941,
        month = sep,
       volume = {94},
        pages = {381},
          doi = {10.1086/144342},
       adsurl = {https://ui.adsabs.harvard.edu/abs/1941ApJ....94..381D}
}

@ARTICLE{Dunham1937,
       author = {{Dunham}, Jr., T.},
        title = "{Interstellar Neutral Potassium and Neutral Calcium}",
      journal = {\pasp},
         year = 1937,
        month = feb,
       volume = {49},
       number = {287},
        pages = {26-28},
          doi = {10.1086/124759},
       adsurl = {https://ui.adsabs.harvard.edu/abs/1937PASP...49...26D}
}

@article{Herbst2020,
author = {Herbst, Eric and Vidali, Gianfranco and Ceccarelli, Cecilia},
title = {Complex Organic Molecules (COMs) in Star-Forming Regions: A Virtual Special Issue},
journal = {ACS Earth and Space Chemistry},
volume = {4},
number = {4},
pages = {488-490},
year = {2020},
doi = {10.1021/acsearthspacechem.0c00043},
URL = {https://doi.org/10.1021/acsearthspacechem.0c00043},
eprint = {https://doi.org/10.1021/acsearthspacechem.0c00043}
}

@ARTICLE{Jimenez-Serra2025,
       author = {{Jimenez-Serra}, Izaskun and {Codella}, Claudio and {Belloche}, Arnaud},
        title = "{Observations of complex organic molecules in the gas phase of the interstellar medium}",
      journal = {arXiv e-prints},
         year = 2025,
        month = mar,
          eid = {arXiv:2503.17104},
        pages = {arXiv:2503.17104},
          doi = {10.48550/arXiv.2503.17104},
archivePrefix = {arXiv},
       eprint = {2503.17104},
 primaryClass = {astro-ph.GA},
       adsurl = {https://ui.adsabs.harvard.edu/abs/2025arXiv250317104J}
}

@ARTICLE{Martin2008,
       author = {{Mart{\'\i}n}, Sergio and {Requena-Torres}, M.~A. and {Mart{\'\i}n-Pintado}, J. and {Mauersberger}, R.},
        title = "{Tracing Shocks and Photodissociation in the Galactic Center Region}",
      journal = {\apj},
         year = 2008,
        month = may,
       volume = {678},
       number = {1},
        pages = {245-254},
          doi = {10.1086/533409},
archivePrefix = {arXiv},
       eprint = {0801.3614},
 primaryClass = {astro-ph},
       adsurl = {https://ui.adsabs.harvard.edu/abs/2008ApJ...678..245M}
}

@ARTICLE{Martin2021,
       author = {{Mart{\'\i}n}, S. and {Mangum}, J.~G. and {Harada}, N. and {Costagliola}, F. and {Sakamoto}, K. and {Muller}, S. and {Aladro}, R. and {Tanaka}, K. and {Yoshimura}, Y. and {Nakanishi}, K. and {Herrero-Illana}, R. and {M{\"u}hle}, S. and {Aalto}, S. and {Behrens}, E. and {Colzi}, L. and {Emig}, K.~L. and {Fuller}, G.~A. and {Garc{\'\i}a-Burillo}, S. and {Greve}, T.~R. and {Henkel}, C. and {Holdship}, J. and {Humire}, P. and {Hunt}, L. and {Izumi}, T. and {Kohno}, K. and {K{\"o}nig}, S. and {Meier}, D.~S. and {Nakajima}, T. and {Nishimura}, Y. and {Padovani}, M. and {Rivilla}, V.~M. and {Takano}, S. and {van der Werf}, P.~P. and {Viti}, S. and {Yan}, Y.~T.},
        title = "{ALCHEMI, an ALMA Comprehensive High-resolution Extragalactic Molecular Inventory. Survey presentation and first results from the ACA array}",
      journal = {\aap},
         year = 2021,
        month = dec,
       volume = {656},
          eid = {A46},
        pages = {A46},
          doi = {10.1051/0004-6361/202141567},
archivePrefix = {arXiv},
       eprint = {2109.08638},
 primaryClass = {astro-ph.GA},
       adsurl = {https://ui.adsabs.harvard.edu/abs/2021A&A...656A..46M}
}

@ARTICLE{McKellar1940,
       author = {{McKellar}, A.},
        title = "{Evidence for the Molecular Origin of Some Hitherto Unidentified Interstellar Lines}",
      journal = {PASP},
         year = 1940,
        month = jun,
       volume = {52},
       number = {307},
        pages = {187},
          doi = {10.1086/125159},
       adsurl = {https://ui.adsabs.harvard.edu/abs/1940PASP...52..187M}
}

@ARTICLE{Rivilla2019,
       author = {{Rivilla}, V.~M. and {Mart{\'\i}n-Pintado}, J. and {Jim{\'e}nez-Serra}, I. and {Zeng}, S. and {Mart{\'\i}n}, S. and {Armijos-Abenda{\~n}o}, J. and {Requena-Torres}, M.~A. and {Aladro}, R. and {Riquelme}, D.},
        title = "{Abundant Z-cyanomethanimine in the interstellar medium: paving the way to the synthesis of adenine}",
      journal = {\mnras},
         year = 2019,
        month = feb,
       volume = {483},
       number = {1},
        pages = {L114-L119},
          doi = {10.1093/mnrasl/sly228},
archivePrefix = {arXiv},
       eprint = {1811.12862},
 primaryClass = {astro-ph.GA},
       adsurl = {https://ui.adsabs.harvard.edu/abs/2019MNRAS.483L.114R}
}

@ARTICLE{Rivilla2020,
       author = {{Rivilla}, V{\'\i}ctor M. and {Mart{\'\i}n-Pintado}, Jes{\'u}s and {Jim{\'e}nez-Serra}, Izaskun and {Mart{\'\i}n}, Sergio and {Rodr{\'\i}guez-Almeida}, Lucas F. and {Requena-Torres}, Miguel A. and {Rico-Villas}, Fernando and {Zeng}, Shaoshan and {Briones}, Carlos},
        title = "{Prebiotic Precursors of the Primordial RNA World in Space: Detection of NH$_{2}$OH}",
      journal = {\apjl},
         year = 2020,
        month = aug,
       volume = {899},
       number = {2},
          eid = {L28},
        pages = {L28},
          doi = {10.3847/2041-8213/abac55},
archivePrefix = {arXiv},
       eprint = {2008.00228},
 primaryClass = {astro-ph.GA},
       adsurl = {https://ui.adsabs.harvard.edu/abs/2020ApJ...899L..28R}
}

@ARTICLE{Rivilla2021,
       author = {{Rivilla}, V{\'\i}ctor M. and {Jim{\'e}nez-Serra}, Izaskun and {Mart{\'\i}n-Pintado}, Jes{\'u}s and {Briones}, Carlos and {Rodr{\'\i}guez-Almeida}, Lucas F. and {Rico-Villas}, Fernando and {Tercero}, Bel{\'e}n and {Zeng}, Shaoshan and {Colzi}, Laura and {de Vicente}, Pablo and {Mart{\'\i}n}, Sergio and {Requena-Torres}, Miguel A.},
        title = "{Discovery in space of ethanolamine, the simplest phospholipid head group}",
      journal = {Proceedings of the National Academy of Science},
         year = 2021,
        month = jun,
       volume = {118},
       number = {22},
          eid = {e2101314118},
        pages = {e2101314118},
          doi = {10.1073/pnas.2101314118},
archivePrefix = {arXiv},
       eprint = {2105.11141},
 primaryClass = {astro-ph.GA},
       adsurl = {https://ui.adsabs.harvard.edu/abs/2021PNAS..11801314R}
}

@ARTICLE{Rivilla2023,
       author = {{Rivilla}, V{\'\i}ctor M. and {Sanz-Novo}, Miguel and {Jim{\'e}nez-Serra}, Izaskun and {Mart{\'\i}n-Pintado}, Jes{\'u}s and {Colzi}, Laura and {Zeng}, Shaoshan and {Meg{\'\i}as}, Andr{\'e}s and {L{\'o}pez-Gallifa}, {\'A}lvaro and {Mart{\'\i}nez-Henares}, Antonio and {Massalkhi}, Sarah and {Tercero}, Bel{\'e}n and {de Vicente}, Pablo and {Mart{\'\i}n}, Sergio and {San Andr{\'e}s}, David and {Requena-Torres}, Miguel A. and {Alonso}, Jos{\'e} Luis},
        title = "{First Glycine Isomer Detected in the Interstellar Medium: Glycolamide (NH$_{2}$C(O)CH$_{2}$OH)}",
      journal = {\apjl},
         year = 2023,
        month = aug,
       volume = {953},
       number = {2},
          eid = {L20},
        pages = {L20},
          doi = {10.3847/2041-8213/ace977},
archivePrefix = {arXiv},
       eprint = {2307.11507},
 primaryClass = {astro-ph.GA},
       adsurl = {https://ui.adsabs.harvard.edu/abs/2023ApJ...953L..20R}
}

@ARTICLE{Rivilla2026,
       author = {{Rivilla}, V.~M. and {San Andr{\'e}s}, D. and {Sanz-Novo}, M. and {Colzi}, L. and {Jim{\'e}nez-Serra}, I. and {L{\'o}pez-Gallifa}, A. and {Mart{\'\i}nez-Henares}, A. and {Meg{\'\i}as}, A. and {Mart{\'\i}n}, S. and {Tercero}, B. and {Zeng}, S. and {Loreau}, J. and {Ben Khalifa}, M. and {Requena-Torres}, M.~A. and {de Vicente}, P.},
        title = "{Aromatic rings in the Central Molecular Zone: Benzonitrile}",
      journal = {arXiv e-prints},
         year = 2026,
        month = apr,
          eid = {arXiv:2604.24510},
        pages = {arXiv:2604.24510},
          doi = {10.48550/arXiv.2604.24510},
archivePrefix = {arXiv},
       eprint = {2604.24510},
 primaryClass = {astro-ph.GA},
       adsurl = {https://ui.adsabs.harvard.edu/abs/2026arXiv260424510R}
}

@ARTICLE{SanAndres2023,
       author = {{San Andr{\'e}s}, D. and {Colzi}, L. and {Rivilla}, V.~M. and {Garc{\'\i}a de la Concepci{\'o}n}, J. and {Melosso}, M. and {Mart{\'\i}n-Pintado}, J. and {Jim{\'e}nez-Serra}, I. and {Zeng}, S. and {Mart{\'\i}n}, S. and {Requena-Torres}, M.~A.},
        title = "{H$_{2}$CN/H$_{2}$NC abundance ratio: a new potential temperature tracer for the interstellar medium}",
      journal = {\mnras},
         year = 2023,
        month = aug,
       volume = {523},
       number = {3},
        pages = {3239-3250},
          doi = {10.1093/mnras/stad1385},
archivePrefix = {arXiv},
       eprint = {2305.04611},
 primaryClass = {astro-ph.GA},
       adsurl = {https://ui.adsabs.harvard.edu/abs/2023MNRAS.523.3239S}
}

@ARTICLE{SanAndres2024,
       author = {{San Andr{\'e}s}, David and {Rivilla}, V{\'\i}ctor M. and {Colzi}, Laura and {Jim{\'e}nez-Serra}, Izaskun and {Mart{\'\i}n-Pintado}, Jes{\'u}s and {Meg{\'\i}as}, Andr{\'e}s and {L{\'o}pez-Gallifa}, {\'A}lvaro and {Mart{\'\i}nez-Henares}, Antonio and {Massalkhi}, Sarah and {Zeng}, Shaoshan and {Sanz-Novo}, Miguel and {Tercero}, Bel{\'e}n and {de Vicente}, Pablo and {Mart{\'\i}n}, Sergio and {Requena Torres}, Miguel Angel and {Molpeceres}, Germ{\'a}n and {Garc{\'\i}a de la Concepci{\'o}n}, Juan},
        title = "{First Detection in Space of the High-energy Isomer of Cyanomethanimine: H$_{2}$CNCN}",
      journal = {\apj},
         year = 2024,
        month = may,
       volume = {967},
       number = {1},
          eid = {39},
        pages = {39},
          doi = {10.3847/1538-4357/ad3af3},
archivePrefix = {arXiv},
       eprint = {2404.03334},
 primaryClass = {astro-ph.GA},
       adsurl = {https://ui.adsabs.harvard.edu/abs/2024ApJ...967...39S}
}

@ARTICLE{SanAndres2026_G0633-I,
       author = {{San Andr{\'e}s}, D. and {Colzi}, L. and {Rivilla}, V.~M. and {Sanz-Novo}, M. and {Mart{\'\i}n}, S. and {Jim{\'e}nez-Serra}, I. and {Zeng}, S.},
        title = "{The Galactic Centre G+0.633-0.0604 molecular cloud: a new astrochemical gold mine. I. Gas physical properties}",
      journal = {arXiv e-prints},
         year = 2026,
        month = jul,
          eid = {arXiv:2607.01481},
        pages = {arXiv:2607.01481},
          doi = {10.48550/arXiv.2607.01481},
archivePrefix = {arXiv},
       eprint = {2607.01481},
 primaryClass = {astro-ph.GA},
       adsurl = {https://ui.adsabs.harvard.edu/abs/2026arXiv260701481S}
}

@ARTICLE{SanAndres2026_G0633-II,
       author = {{San Andr{\'e}s}, D. and {Rivilla}, V.~M. and {Colzi}, L. and {author}, A. and {author}, B. and {authors}, C.},
      journal = {\aap},
         year = 2026,
       volume = {submitted},
          eid = {submitted},
}

@ARTICLE{Sanz-Novo2023,
       author = {{Sanz-Novo}, Miguel and {Rivilla}, V{\'\i}ctor M. and {Jim{\'e}nez-Serra}, Izaskun and {Mart{\'\i}n-Pintado}, Jes{\'u}s and {Colzi}, Laura and {Zeng}, Shaoshan and {Meg{\'\i}as}, Andr{\'e}s and {L{\'o}pez-Gallifa}, {\'A}lvaro and {Mart{\'\i}nez-Henares}, Antonio and {Massalkhi}, Sarah and {Tercero}, Bel{\'e}n and {de Vicente}, Pablo and {Mart{\'\i}n}, Sergio and {San Andr{\'e}s}, David and {Requena-Torres}, Miguel A.},
        title = "{Discovery of the Elusive Carbonic Acid (HOCOOH) in Space}",
      journal = {\apj},
         year = 2023,
        month = sep,
       volume = {954},
       number = {1},
          eid = {3},
        pages = {3},
          doi = {10.3847/1538-4357/ace523},
archivePrefix = {arXiv},
       eprint = {2307.08644},
 primaryClass = {astro-ph.GA},
       adsurl = {https://ui.adsabs.harvard.edu/abs/2023ApJ...954....3S}
}

@ARTICLE{Sanz-Novo2026,
       author = {{Sanz-Novo}, M. and {Rivilla}, V.~M. and {Jim{\'e}nez-Serra}, I. and {Colzi}, L. and {Zeng}, S. and {Meg{\'\i}as}, A. and {San Andr{\'e}s}, D. and {L{\'o}pez-Gallifa}, {\'A}. and {Mart{\'\i}nez-Henares}, A. and {Fried}, Z.~T.~P. and {McGuire}, B.~A. and {Mart{\'\i}n}, S. and {Requena-Torres}, M.~A. and {Tercero}, B. and {de Vicente}, P. and {Kolesnikov{\'a}}, L. and {Alonso}, E.~R. and {Cocinero}, E.~J. and {Guillemin}, J.~C. and {Kleiner}, I.},
        title = "{Expanding the C$_{3}$H$_{6}$O$_{2}$ isomeric interstellar inventory: Discovery of lactaldehyde and methoxyacetaldehyde in G+0.693-0.027}",
      journal = {\aap},
         year = 2026,
        month = feb,
       volume = {706},
          eid = {A316},
        pages = {A316},
          doi = {10.1051/0004-6361/202558316},
archivePrefix = {arXiv},
       eprint = {2601.07365},
 primaryClass = {astro-ph.GA},
       adsurl = {https://ui.adsabs.harvard.edu/abs/2026A&A...706A.316S}
}

@ARTICLE{Swings&Rosenfeld1937,
       author = {{Swings}, P. and {Rosenfeld}, L.},
        title = "{Considerations Regarding Interstellar Molecules}",
      journal = {ApJ},
         year = 1937,
        month = nov,
       volume = {86},
        pages = {483-486},
          doi = {10.1086/143880},
       adsurl = {https://ui.adsabs.harvard.edu/abs/1937ApJ....86..483S}
}

@ARTICLE{Weinreb1963,
       author = {{Weinreb}, S. and {Barrett}, A.~H. and {Meeks}, M.~L. and {Henry}, J.~C.},
        title = "{Radio Observations of OH in the Interstellar Medium}",
      journal = {Nature},
         year = 1963,
        month = nov,
       volume = {200},
       number = {4909},
        pages = {829-831},
          doi = {10.1038/200829a0},
       adsurl = {https://ui.adsabs.harvard.edu/abs/1963Natur.200..829W}
}
\bibliographystyle{aa}

\end{document}